\documentclass[twocolumn]{aastex63}

\newcommand{\gr}{$\gamma$-ray}
\newcommand\fermi{{\it Fermi}}

\begin{document}

\title{A Possible Gamma-Ray Enhancement Event in Tycho's Supernova Remnant}

\correspondingauthor{Zhongxiang Wang \& Xiao Zhang}
\email{wangzx20@ynu.edu.cn, xiaozhang@nju.edu.cn}

\author{Yi Xing}
\affiliation{Shanghai Astronomical Observatory, Chinese Academy of Sciences,
80 Nandan Road, Shanghai 200030, China}

\author{Zhongxiang Wang}
\affiliation{Department of Astronomy, Yunnan University, Kunming 650091, China}
\affiliation{Shanghai Astronomical Observatory, Chinese Academy of Sciences, 
80 Nandan Road, Shanghai 200030, China}

\author{Xiao Zhang}
\affiliation{School of Astronomy \& Space Science, Nanjing University, 
163 Xinlin Avenue, Nanjing 210023, China}
\affiliation{Key Laboratory of Modern Astronomy and Astrophysics, Nanjing University, Ministry of Education, China}

\author{Yang Chen}
\affiliation{School of Astronomy \& Space Science, Nanjing University, 
163 Xinlin Avenue, Nanjing 210023, China}
\affiliation{Key Laboratory of Modern Astronomy and Astrophysics, Nanjing University, Ministry of Education, China}

\begin{abstract}
	We report a possible $\gamma$-ray enhancement event detected from
Tycho's supernova remnant (SNR), the outcome of a type Ia supernova
explosion that occurred in year 1572. The event lasted for 1.5 years and showed
a factor of 3.6 flux increase mainly in the energy range of 4--100~GeV, 
	while notably accompanied with two 478\,GeV photons.
Several young SNRs (including Tycho's SNR)
were previously found to show peculiar X-ray structures with flux variations in
one- or several-year timescales, such an event at $\gamma$-ray energies
	is for the first time seen. The year-long timescale of the 
	event suggests a synchrotron radiation process, but the
	hard $\gamma$-ray emission 
	requires extreme conditions of either
ultra-high energies for the electrons upto $\sim$10~PeV
(well above the cosmic-ray ``knee" energy) or high inhomogeneity of 
	the magnetic field in the SNR.
This event in Tycho's SNR is likely analogous to the $\gamma$-ray flares
observed in the Crab nebula, the comparably short timescales
of them both requiring a synchrotron process, and similar
magnetohydrodynamic processes such as magnetic reconnection would be
at work as well in the SNR to accelerate particles to ultra-relativistic
energies.
	The event, if confirmed,  helps reveal the more complicated 
	side of the physical processes that can occur in young SNRs.

\end{abstract}


\section{Introduction}

In 1572, a supernova explosion was recorded by Tycho Brahe, of which
the remnant is referred to as Tycho's supernova remnant (SNR; hereafter Tycho).
This supernova is believed to have a type Ia origin \citep{baa45,rui+04,kra+08},
i.e., the thermonuclear explosion of a CO white dwarf whose mass has reached
close to the Chandrasekhar limit ($\sim$1.4$\,M_{\odot}$) through mass
accretion \citep{lm18}.
Being young and one of a few known type Ia SNRs in our Galaxy, Tycho has been
well studied. While the distance is uncertain, estimated
to be in a range of 2--5~kpc (\citealt{hay+10,tl11}; we adopt 2.5~kpc in this
work), the SNR has
a shell-like shape with a size of $\sim$8~arcmin in diameter in high-resolution
images of multiple
wavelengths \citep{rey+97,kot+06,hwa+02,wil+13,lop+15}. The shell is expanding,
of which the ejecta was found to have an expansion velocity of 
$\sim$4700~km~s$^{-1}$ \citep{hay+10}, carrying much of
the energy released in the supernova explosion. At the shell front,
acceleration of high-energy electrons is suggested to be
seen \citep{bam+05},
which is considered as evidence for cosmic ray production.

Through the diffusive shock acceleration process \citep{bel78,bo78,dru83},
cosmic rays below the so-called ``knee" ($\sim$3~PeV) are widely believed to
arise primarily from SNRs in our Galaxy. High-energy \gr\ studies of SNRs
can thus  help find possible evidence on
the origin of cosmic rays \citep{dav94}. Specifically for Tycho, its
high-energy and very-high-energy (VHE) \gr\ emission has been
detected \citep{acc+11,gio+12,arc+17}. With the focus on the
\gr\ detection, models to explain the observed emission from Tycho
have been proposed \citep{ad12,mc12,bkv13,zha+13,sla+14}. Most of them indicate
that the hadronic process, the collisions of high-energy hadrons (protons and
ions), is
responsible for the \gr\ production. X-ray studies of Tycho have also revealed
intriguing features, in particular the stripe-like structures that are
suggested as
evidence of particles accelerated to PeV energies \citep{eri+11}. Very recent
results from analysis of the long-term data show that these X-ray stripes in
the southwestern region of
the SNR were variable, and are interpreted as the indication of the magnetic
field amplification and/or significant magnetic turbulence
changes \citep{oku+20,mat+20}. In addition, rapid deceleration of the expanding
shell was found in year 2009--2015 in particular \citep{tan+20}.

Likely related to these features and their implications, here we report
a possible \gr\ enhancement phenomenon seen in this SNR. 
In section~\ref{sec:obs},
detailed analysis of the data obtained with the Large Area Telescope (LAT)
onboard {\it Fermi Gamma-ray Space Telescope (Fermi)} is presented. In
section~\ref{sec:res}, the results from the analysis are summarized and
the implications of the results are discussed.

\section{LAT Data Analysis}
\label{sec:obs}

\subsection{Analysis of the 11.6-yr \fermi\ LAT data}
We selected LAT events from the latest Pass 8 database in an energy range of
60 MeV to 500~GeV during the time period from 2008-08-04 15:43:36 (UTC) to 
2020-03-05 01:16:35 (UTC).  The region of interest (RoI) is 
$20^{\circ}\times 20^{\circ}$, centered at the position of Tycho.
Only the events with quality flags of `good' and zenith angles smaller than 
90~degrees were used; the latter is to prevent 
the Earth's limb contamination. Both of these are recommended by the LAT 
team\footnote{\footnotesize http://fermi.gsfc.nasa.gov/ssc/data/analysis/scitools/}.

We constructed a source model based on the \fermi\ LAT 10-year source catalog
\citep{bal+20}, by including all the sources within 20 degrees away from 
Tycho. The spectral forms and parameters of these sources are provided in 
the catalog. In our analysis, we fixed the spectral parameters of the sources 
5~degrees away from the target at the values given in the catalog, and set 
those of the sources within 5 degrees as free parameters. We included 
the background Galactic and extragalactic diffuse 
spectral models (gll\_iem\_v07.fits and iso\_P8R3\_SOURCE\_V2\_v1.txt
respectively) in the source model, and their normalizations were set 
as free parameters. The emission from Tycho was described with a simple power law $\sim E^{-\Gamma}$, where $E$ is
the photon energy and $\Gamma$ is the photon spectral index.
\begin{figure}
\begin{center}
\includegraphics[width=0.77\linewidth]{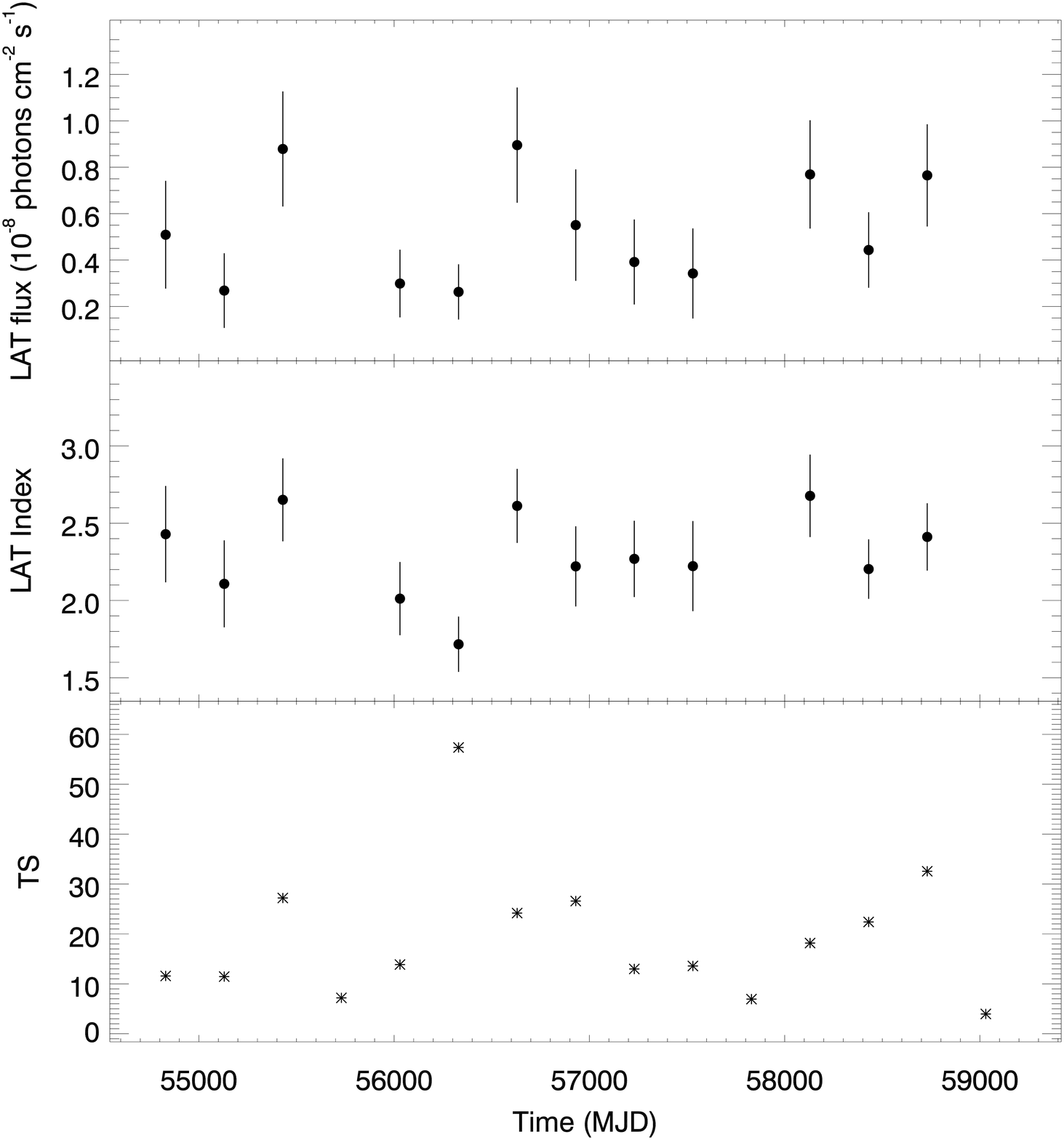}
\end{center}
	\caption{300-day binned 
	light curve of Tycho ({\it top} panel), in which data points
	with TS$\geq$9 are shown.
	The corresponding power-law index and TS are shown in the {\it middle}
	and {\it bottom} panels respectively. 
	}
\label{fig:lc300}
\end{figure}

We performed the binned likelihood analysis to the LAT data in
the energy range of 0.3--500~GeV during the whole selected
time period. In the energy range $<$0.3~GeV, the instrument
response function of LAT has relatively large uncertainties
and the background emission along the Galactic plane is strong \citep{4fgl20}.
A photon index of $\Gamma= 2.23\pm$0.07 and a 0.3--500~GeV flux of
$F_{0.3-500}= 4.3\pm 0.5\times 10^{-9}$ photons~s$^{-1}$\,cm$^{-2}$ were
obtained (Table~\ref{tab:gam}), which are consistent with those given in the \fermi\ LAT 10-year
source catalog. The Test Statistic (TS) value obtained is 245. 

\subsection{Variability and spectral analysis}

To check the long-term emission properties of Tycho, we first extracted the \gr\ light curves of the source.
Because Tycho is faint, we initially constructed 0.3--500\,GeV light 
curves of 100, 200, 300, 400, and 500-day bins. The 200- and 300-day binned 
light
curves relatively well show possible variations, as the time bins 
can avoid
to either have many flux upper limits (when too small) or smear out possible
changes (when too large). For example, there are only three data points in
the 300-day binned light curve (Figure~\ref{fig:lc300}) with TS$<$9 (resulting in flux upper limits in
this case), and most of them are in the range of 10$<$TS$<$30.
Also notably the data point at MJD 56330 has TS$\sim 60$ and a hard 
spectral index, suggesting possible emission changes. 
\begin{figure*}
\begin{center}
\includegraphics[width=0.47\linewidth]{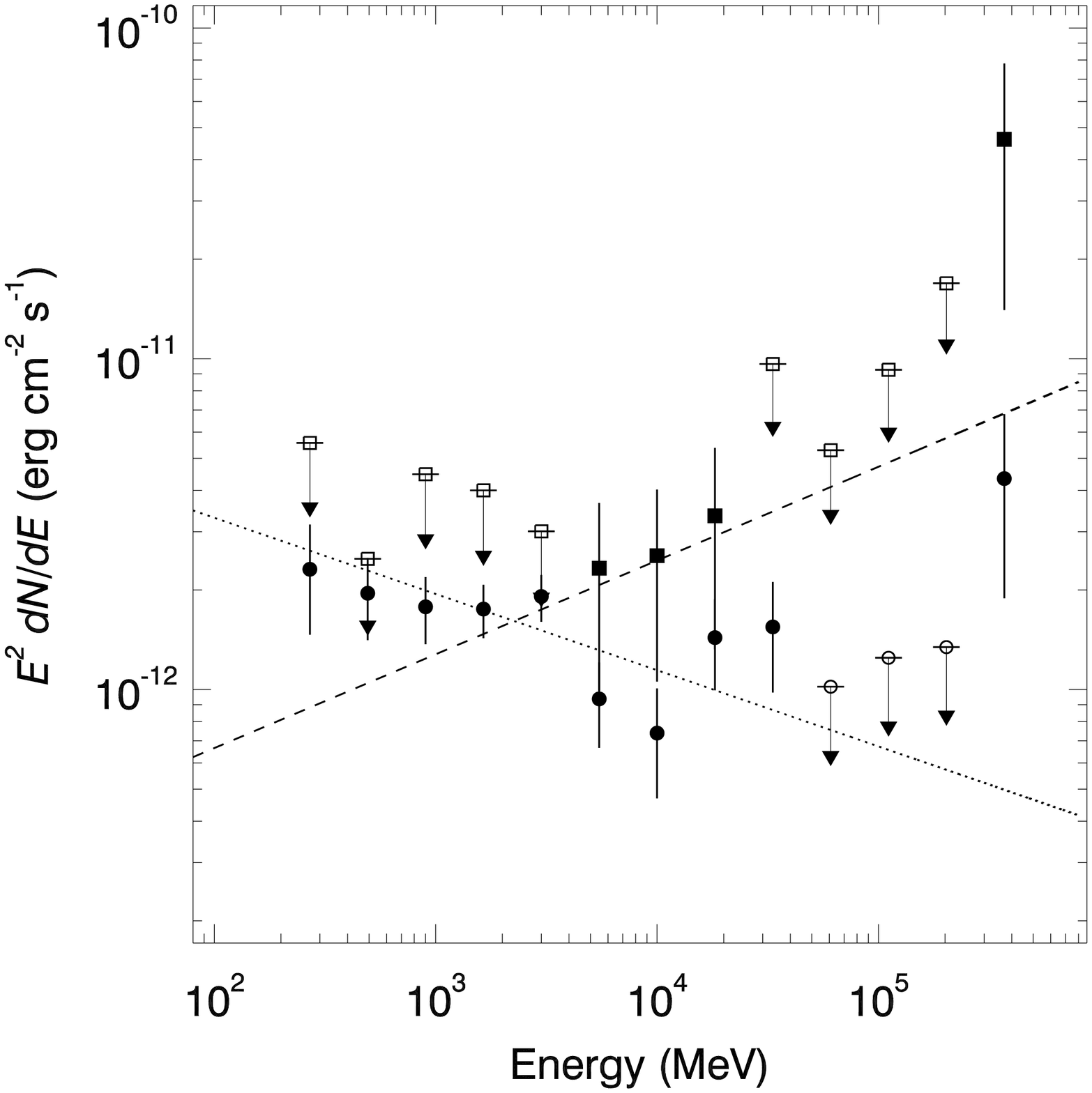}
\includegraphics[width=0.47\linewidth]{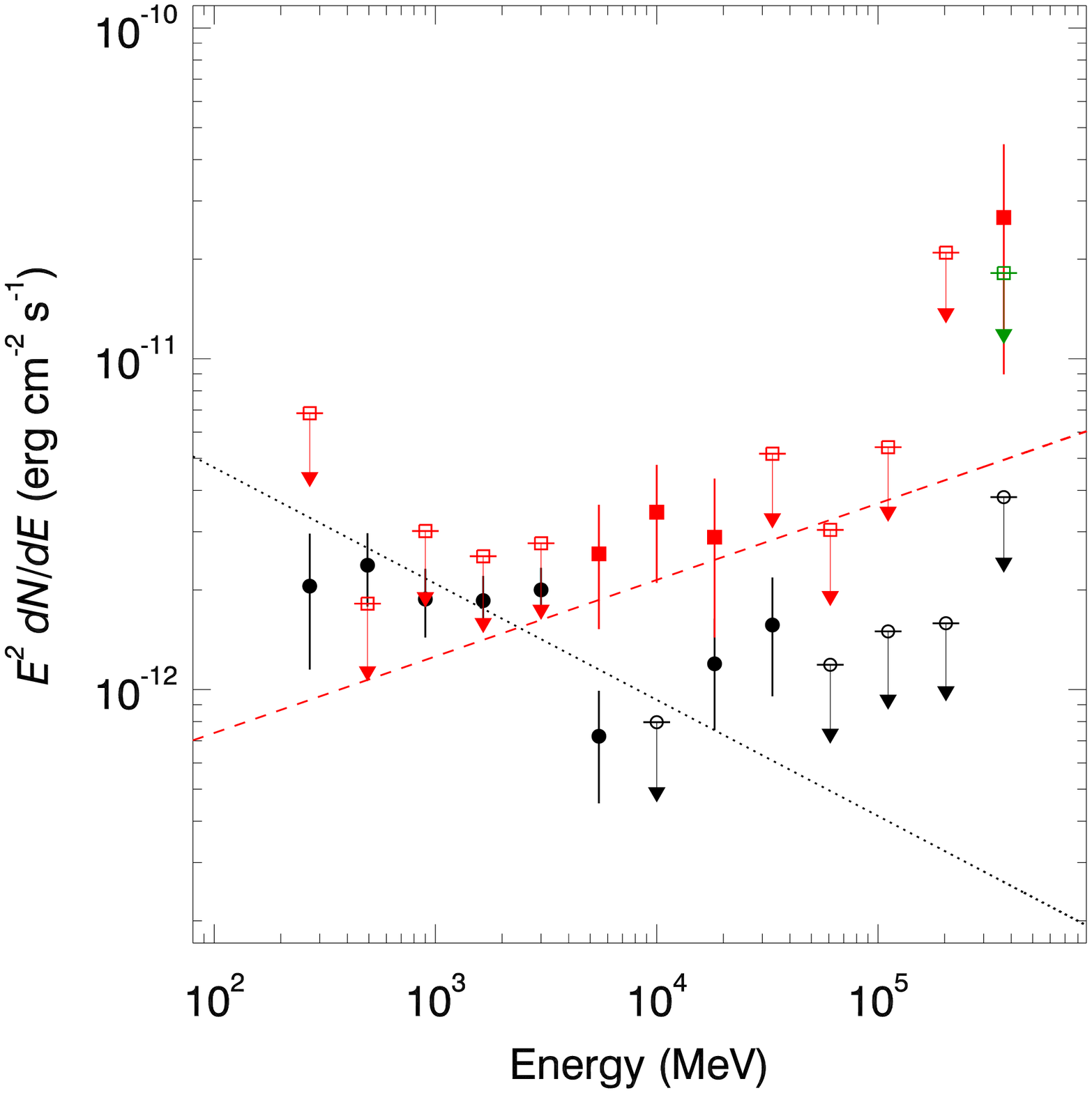}
\end{center}
	\caption{{\it Left:} Spectra of Tycho obtained from the whole data (11.6 yrs; dots)
	and 300-day data in MJD~56180--56480 (squares) in the energy range of 
	0.2--500\,GeV. The corresponding power-law fits obtained from the 
	binned likelihood analysis are shown as dotted and dashed lines 
	respectively. {\it Right:} Spectra of Tycho in the energy range of 0.2--500~GeV in time period
II (red squares) and the remaining time periods (black dots). 
	The red dashed and black dotted lines are
the best-fit power-law models obtained from the binned likelihood analysis
for the two spectra. When we do not include the two $\sim$478~GeV photon
events (see Section~\ref{sec:prob}), the data point in the energy 
range of 273.9--500~GeV is turned to be a flux upper limit (the green square).
	In the both panels, the open signs are the flux upper limits. 
	}
\label{fig:spec}
\end{figure*}

We thus extracted a spectrum from the 300-day data centered at 
MJD~56330 by performing likelihood analysis in 13 evenly divided
energy bands in logarithm in 0.2--500\,GeV energy range. The spectrum of
the same energy bands
from the whole data was also extracted.
The spectral normalizations of the sources within 5 degrees from the target
were set as free parameters, while all the other parameters
of the sources in the source model were fixed at the values obtained from
the above binned likelihood analysis.
For the obtained spectral data points, we kept those with TS greater than
9, and derived 95\% flux upper limits otherwise.
The spectra are shown in the left panel of Figure~\ref{fig:spec}.
The spectral points of the whole data set are given in Table~\ref{tab:spec}.

Comparing the two spectra in Figure~\ref{fig:spec}, possible differences 
are seen in the energy range 
of 4.1--25\,GeV. The three spectral data points 
covering the energy range appear higher in the 300-day spectrum
than those in the whole-data spectrum,
although their uncertainties are large.
It can be noted that the last
data point (at $\sim$370\,GeV) in the two spectra is not an upper limit.
This flux data point is actually due to the two $\sim$478~GeV photons (see
section~\ref{sec:prob}).

\begin{figure}
\begin{center}
\includegraphics[width=0.95\linewidth]{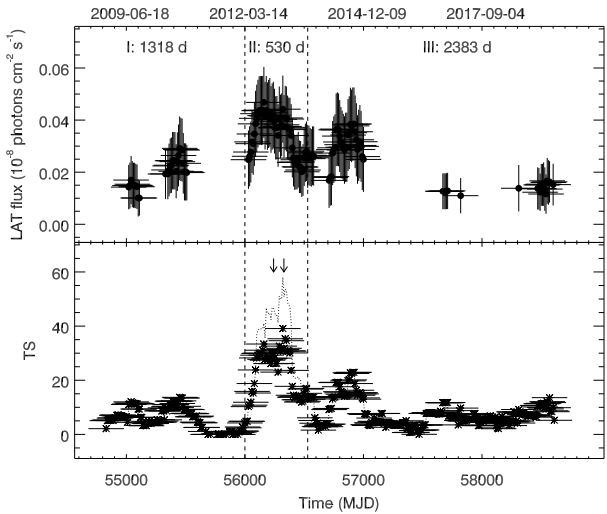}
\end{center}
\caption{
300-day binned light curve ({\it top} panel) and TS curve ({\it bottom}
panel) of Tycho in the energy range of 4.1--100~GeV, where each data point
is constructed by shifting 10 days (instead of 300 days) forward and 
each horizontal bar
indicates the length of 300 days. Only the flux data points with
TS$\geq$9 are shown in the top panel.
The TS curve shows a time period of 530 days for having
TS$\geq 30$, for which we define as time period II (MJD~56000--56530).
When extend the energy range of the data to 500~GeV, which thus include two
$\sim$478~GeV events (with the arrival times indicated by two arrows), the TS
values during time period II are further increased (indicated by the dotted
curve).
\label{fig:lc_4100}}
\end{figure}

The energy range of $\geq$4.1\,GeV was thus chosen for determining 
the time range containing possible flux variations of Tycho.
A 300-day binned smooth light curve was constructed, for which each data point
was shifted by 10 days forward (instead of 300 days). This type of
smooth light curves helps reveal possible fine structures of variations.
From the smooth light curve, we determined the time range of MJD~56000--56530
(defined as time period II)
in which Tycho showed an emission enhancement.
This time period was set by requiring the TS values in time bins greater than 30
(Figure~\ref{fig:lc_4100}).  The time periods
before and after it (1318 and 2383 days respectively) were defined as 
time period I and III respectively.

We then extracted the \gr\ spectra of Tycho in the energy range of 0.2--500 GeV
during the time period II and the remaining time periods.
The spectra are shown in the right panel of Figure~\ref{fig:spec}, and the data points are
provided in Table~\ref{tab:spec}.
We note that during time period II, the last spectral data point,
a flux measurement, is in 273.9--500 GeV band, which includes
the two $\sim$478~GeV events (see
section~\ref{sec:prob}). If we exclude the two events, the data point
turns to be a flux upper limit instead.

\subsection{Energy distribution of photon events}
\label{sec:prob}

During the time period II, there are two events with energies of 478.5 and
478.3~GeV at MJD~56242 and 56329 respectively. The event class 
{\tt Source}, recommended for most analyses with good sensitivity for point
sources, was used in our analysis. At $\sim$500\,GeV, the 68\% containment
angle for the event class is about 0.1 degree. We ran {\tt gtfindsrc} to 
determine the position
of the \gr\ emission in the energy range of $\geq 100$\,GeV during the time 
period II. The position has R.A. = 6\fdg22 and Decl. = 64\fdg13, with an
1$\sigma$ uncertainty of 0\fdg04. Tycho is 0\fdg05
away from the position, within the 2$\sigma$ error circle (consistent with
the containment angle). The two 
$\sim$478\,GeV events are the only high-probability ones in association with
Tycho among detected photons in the high energy band. We note that
the energy resolution at $\sim$500\,GeV is less than 10\%, set as the 68\%
containment half width of the reconstructed incoming photon energy.

In order to understand the association of the two $\sim$478~GeV events with
Tycho and also check if there are other related events with slightly lower 
energies, we calculated the association probabilities of the photon 
events for Tycho. 
We chose a circular region centered at Tycho with 3-degree radius, 
which 
approximately corresponds to the 68\% containment angle of the Point Spread
Function (PSF) of the LAT at 200~MeV\footnote{\footnotesize https://www.slac.stanford.edu/exp/glast/groups/canda/lat\_Performance.htm}.
The association probabilities of the events in this region
for all the sources in our source model 
(i.e., the 20-degree radius RoI centered at Tycho) were calculated,
using the LAT data analysis tool {\tt gtsrcprob}.
The two 478.5 and 478.3~GeV events are highly
associated with Tycho, having the probabilities of 94\% and 98\% coming
from Tycho, respectively.

\begin{figure*}
\begin{center}
\includegraphics[width=0.47\linewidth]{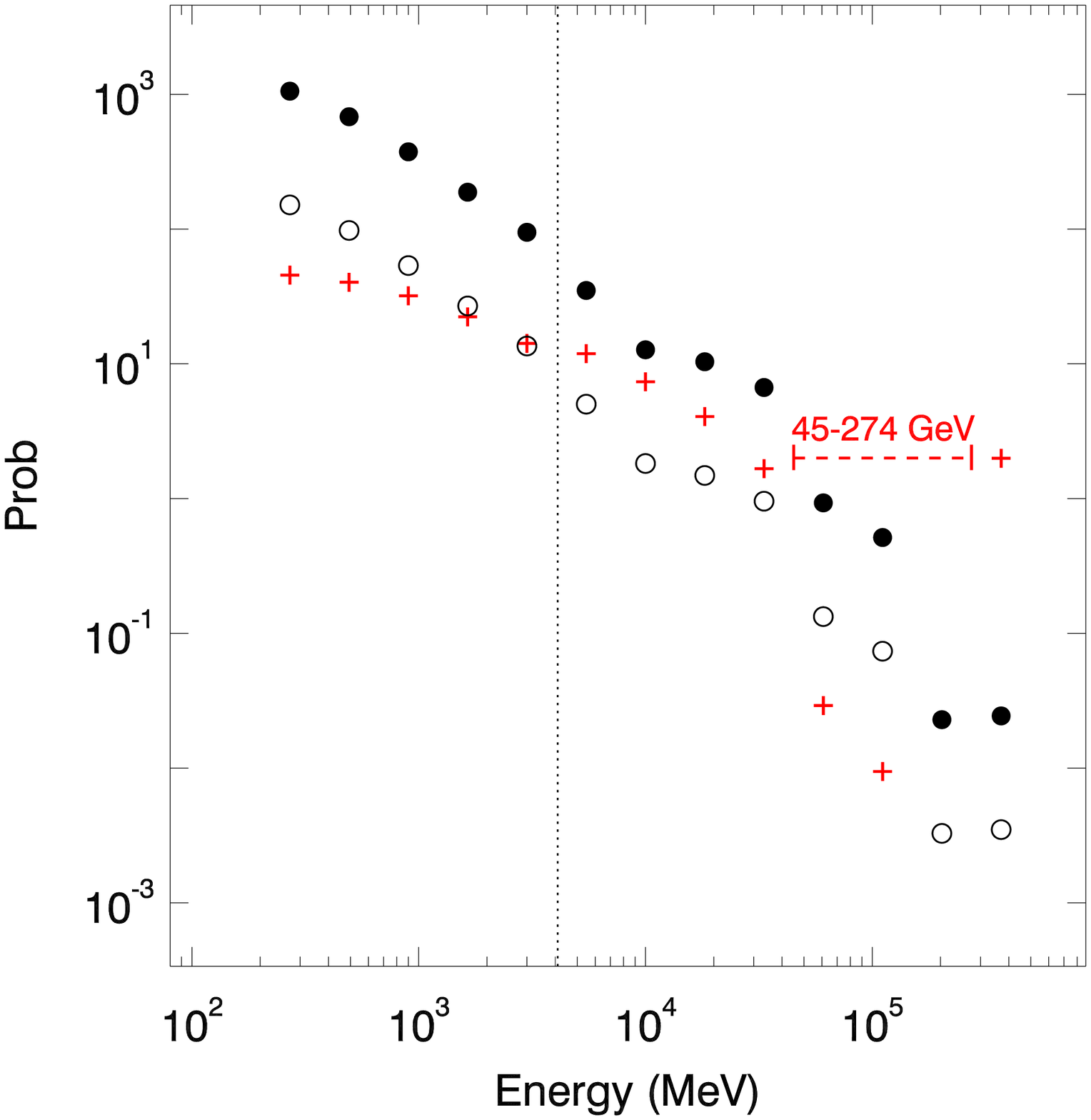}
\includegraphics[width=0.47\linewidth]{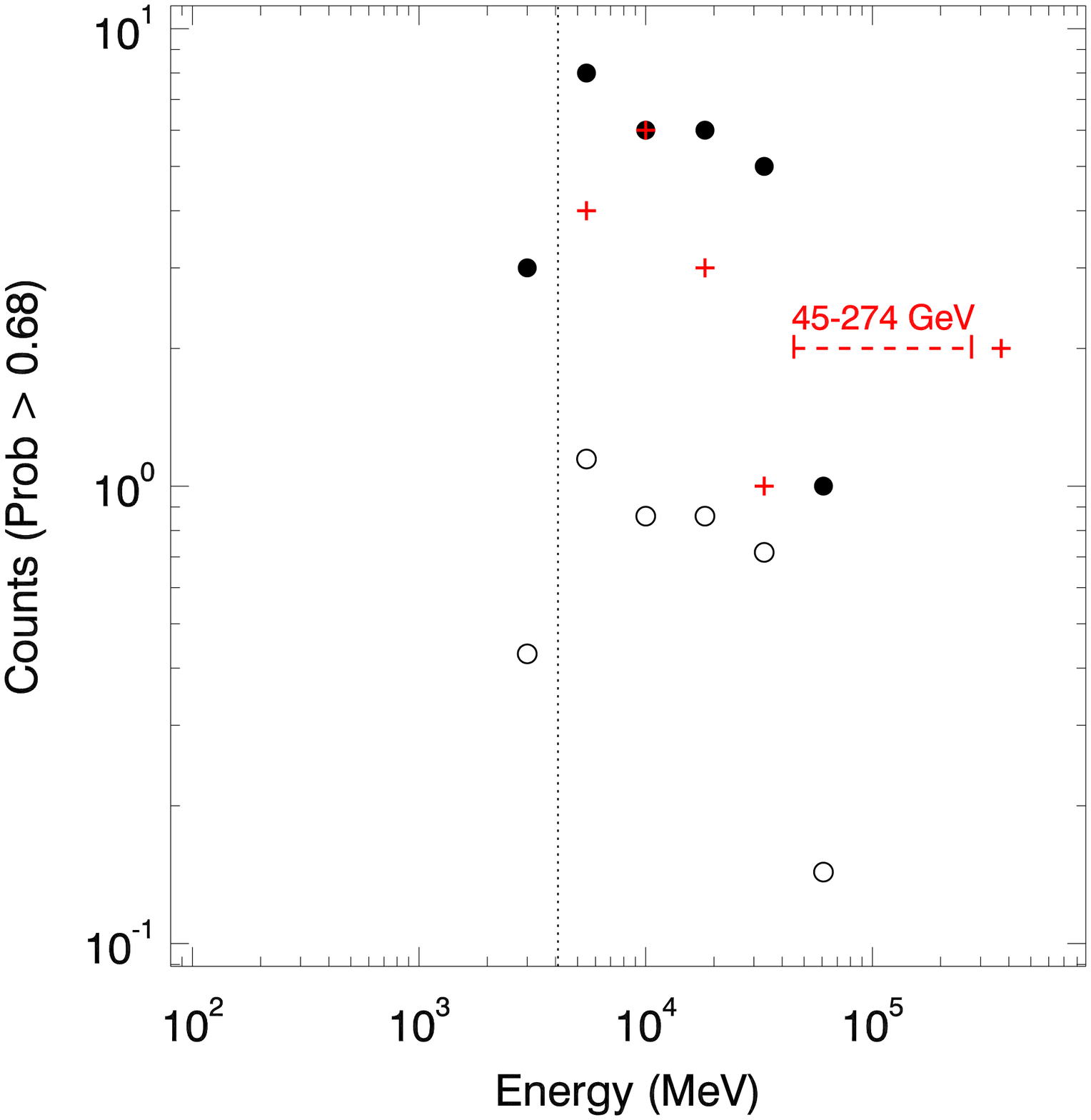}
\end{center}
	\caption{{\it Left:} Added probabilities of photon events for 
	Tycho in the energy range of 0.2--500\,GeV. The black circles are 
	values in
	time period I+III, while the open circles are values scaled to the 
	length of time period II. The red pluses are values in time period II.
	{\it Right:} Photon counts with $\geq$68\% association 
	probabilities for Tycho. The symbols are the same as those in the left 
	panels. Due to high backgrounds, few photons are assigned with 
	high probability values below 4\,GeV (indicated by the vertical 
	dotted line). There is a gap with very low or no photons between
	45--274\,GeV in time period II, indicating that the two 
	$\sim$478\,GeV photons (the last data point) are isolated events.}
\label{fig:prob}
\end{figure*}

We set the energy bins the same as those for the above spectrum
extraction. In each energy bin, the probabilities of the events for
Tycho were added (i.e., the weighted counts from Tycho). In the left panel of Figure~\ref{fig:prob}, 
we show the probability results for time periods I+III 
and II: the values in the former are generally higher than those in the latter,
but when the lengths of the time periods are considered, the values in 
4.1--45\,GeV are higher in the latter than in the former (very similar to 
the spectrum results shown above). In addition, the two $\sim$478\,GeV events
constitutes the last data point with the added probability close to 2. 
It can be noted that between 45--274\,GeV, there is an obvious gap with very low
probability values.

Another figure can be plotted is photon events with high association 
probabilities. We counted the numbers of events in the 3-degree radius region 
with $\geq$68\% association probabilities for Tycho. The same energy bins 
were used. The counts results are shown in the right panel of Figure~\ref{fig:prob}. 
In time period I+III, when we scale its counts with the length ratio between
time period I+III and time period II, the counts are mostly $<$1 in
the energy bins between 2.2--82\,GeV.
On the other hand in time period II,
the counts are mostly $>$1 in the energy bins between 4.1--45\,GeV
and the highest is 6 at 10\,GeV.
Also similarly to the above probability plot, the last data 
point is close to 2 due to the two $\sim$478\,GeV photons and 
the gap between 45--274\,GeV is visible. Thus the two photons were 
isolated events, not coming together with a beam of similarly high energy 
events from Tycho.

\subsection{Analysis of the 4.1--100~GeV LAT data}
As the two  $\sim$478\,GeV events can significantly affect the results such as by
increasing the TS value or changing the spectrum, we
used the energy range of 4.1--100~GeV in our binned
likelihood analysis to avoid inclusion of the two $\sim$478 GeV events.
The obtained 4.1--100~GeV
fluxes are $4.0\pm 0.9\times 10^{-10}$ photons~s$^{-1}$\,cm$^{-2}$
and $1.1\pm 0.2\times 10^{-10}$ photons~s$^{-1}$\,cm$^{-2}$ during time
period II and time period I plus III, respectively (Table~\ref{tab:gam}). The integrated flux in 
the former
is thus higher than that in the latter at a 3$\sigma$ confidence level.
We checked the normalization factors of the Galactic background, the dominant
background for Galactic sources (the Galactic latitude of Tycho is 1.4 degrees),
and they were 1.03$\pm$0.02 and 1.001$\pm$0.008 in the two time periods,
consistent with each other.
For completeness of the results, we also obtained the fluxes in the energy 
range of 0.3--500~GeV
in time period I, II, and III, and the values are also given in Table~\ref{tab:gam}.

\begin{figure*}
\centering
\includegraphics[width=0.39\linewidth]{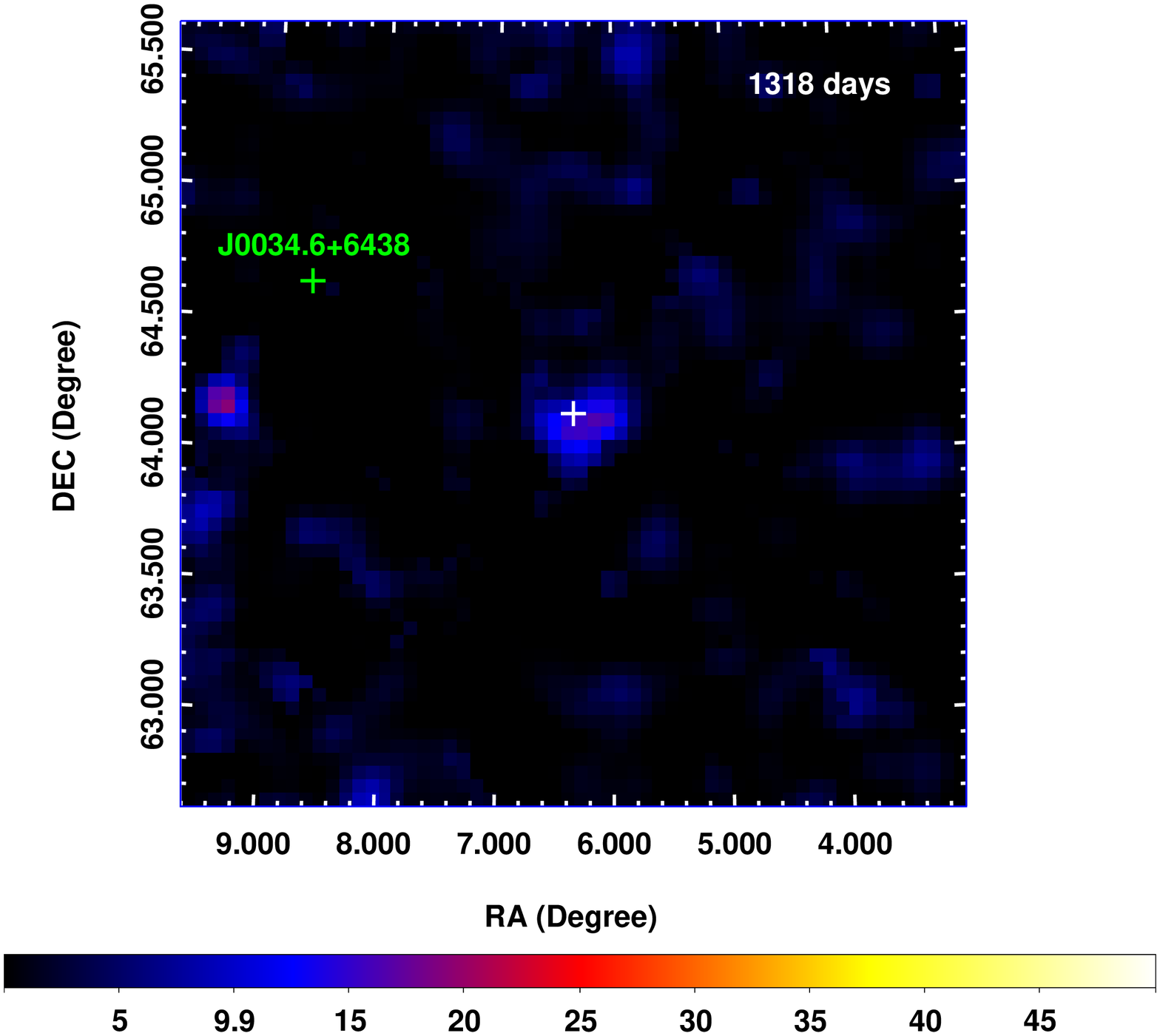}
\includegraphics[width=0.39\linewidth]{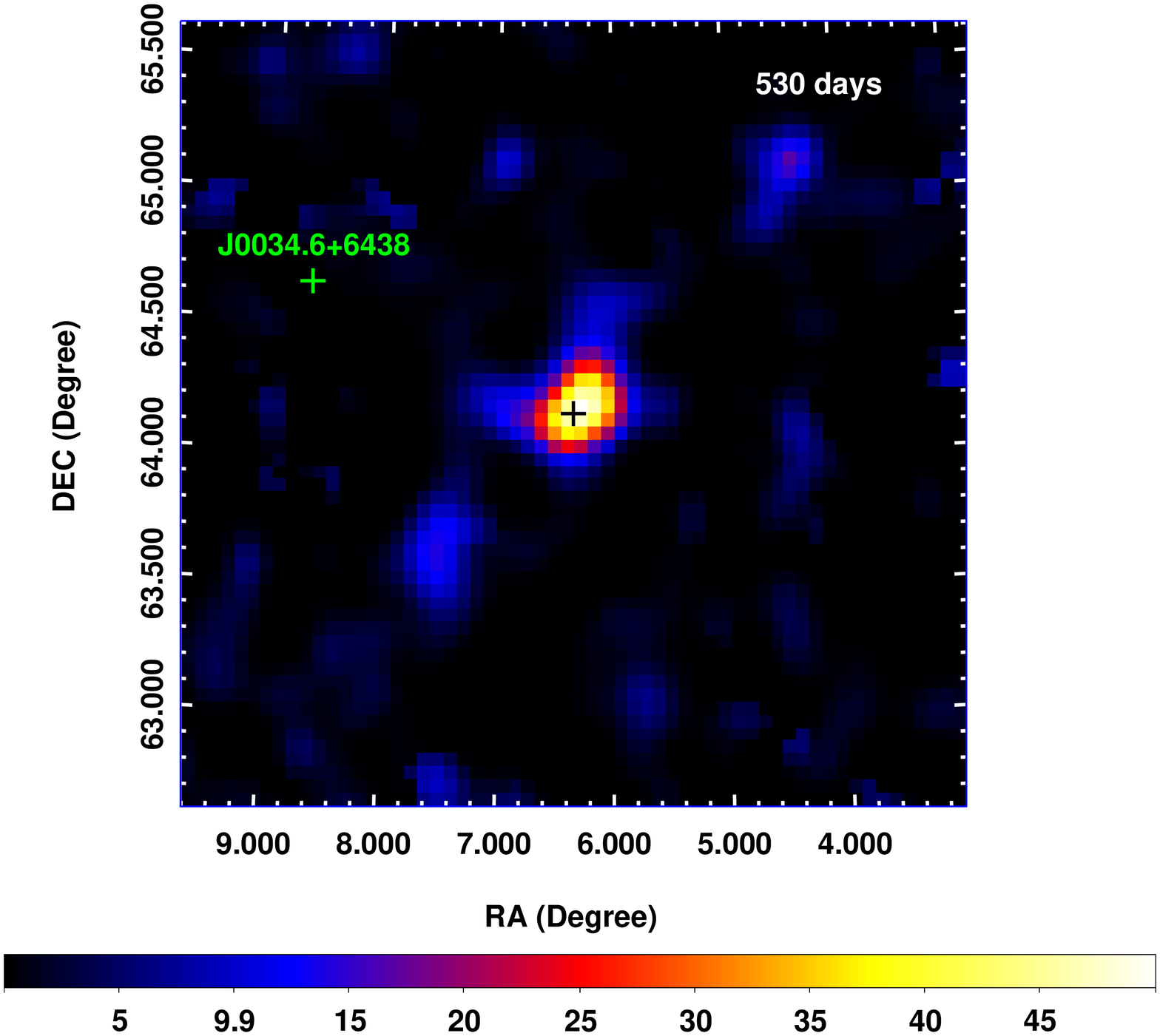}
\includegraphics[width=0.39\linewidth]{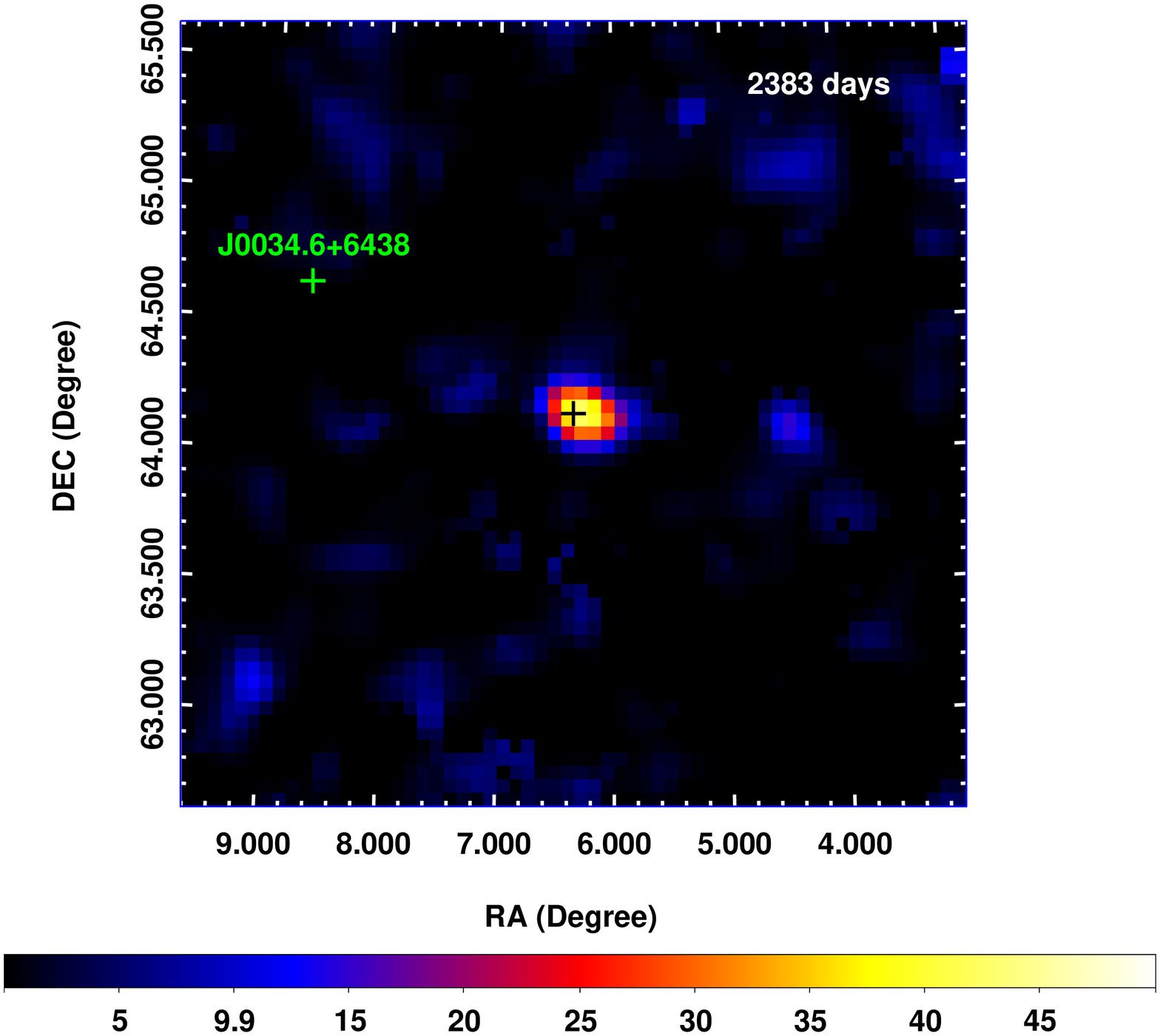}
\caption{TS maps of the $3^{\circ}\times 3^{\circ}$ region centered
at Tycho in the energy range of 4.1--100 GeV from the data of
time period I ({\it top left} panel), time period II ({\it top right} panel),
and time period III ({\it bottom} panel).
The white (or black) plus in each panel marks the position of Tycho, and
the green plus marks a catalog source that is included in the source model
and thus removed when calculating the TS maps.
The image scale of the TS maps is 0.05$^{\circ}$\,pixel$^{-1}$.
\label{fig:ts}
}
\end{figure*}

A TS map helps show if a significant detection (with a high TS value
at a source's position) is due to the presence of the source at the position 
or not.
To carefully examine the detection of the enhanced emission, we 
calculated the 4.1--100 GeV TS maps centered at Tycho in the three time 
periods. 
The three TS maps are shown in Figure~\ref{fig:ts}. At the position of Tycho, 
there is always the point-source emission present without any additional 
emission sources.
The highest TS value of $\simeq$50 was found in time period II, 
comparing to 15 and 41 in time period I and III respectively 
(note that the latter two have time durations 2.5 and 4.5 times longer 
than that of the former).  The results clearly
indicate that there was enhanced emission in time period II.

Also noted is that the \gr\ field around Tycho is rather clean. There is only 
one faint unidentified source (J0034.6+6438; see Figure~\ref{fig:ts})
1.1~degrees away from Tycho.
Its emission
is described with a power law with $\Gamma\simeq$2.43. It is not variable
with a TS value of 80 in the LAT source catalog. Considering that the 95\%
containment angle of the LAT PSF is $\simeq$1\,degree
at 4\,GeV\footnote{\footnotesize https://www.slac.stanford.edu/exp/glast/groups/canda/lat\_Performance.htm}, this source did not likely contaminate the enhancement detection.
The 2$\sigma$ positional 
uncertainty for the \gr\ emission of Tycho is 0.05 degrees. Within
the error region, no blazars (the dominant \gr\ sources in the sky) are known
to be located. Thus no evidence shows that the enhancement would be caused
by a background source.

\section{Results and Discussion}
\label{sec:res}

Using the \fermi\ LAT data, we have conducted detailed analysis for studying
the \gr\ emission of Tycho.
We found that 
the long-term flux and TS values in the energy range of
4.1--100 GeV, obtained from 300-day time bin LAT data, 
effectively demonstrate the enhancement in emission from Tycho 
(Figure~\ref{fig:lc_4100}). During a time period of
530 days ($\simeq$1.5~yr between MJD~56000--56530),
the TS values
are greater than 30, higher than every of those in the remaining time
periods (time periods I and III with lengths of 1318 and 2383 days
respectively).
Interestingly, there are two $\sim$478~GeV photon events received from
the direction of Tycho during time period II, and the association probabilities,
respectively 94\% and 98\%, with Tycho are very high.
Because of the rareness of such 
high-energy photons \citep{2fhl}, the two events can significantly affect
the results. For example, if we extend the energy range to 500~GeV, 
the TS values in time period II are further increased due to 
the two events (Figure~\ref{fig:lc_4100}).

The main discernable difference in emission of Tycho during time 
period II, from that in the remaining time periods (i.e., time period I plus 
III), can be seen to occur in the energy range of 
4--25~GeV (Figure~\ref{fig:spec}).  
The results from the binned likelihood analysis (Table~\ref{tab:gam})
indicate that the emission appears marginally softer
(larger $\Gamma$ value). 
The flux in this time period is
$4.0\pm0.9\times 10^{-10}$\,photons\,cm$^{-2}$\,s$^{-1}$ (in the energy range
of 4.1--100\,GeV), higher by a factor 
of $\sim$3.6 than that in the remaining time periods, which is
$1.1\pm0.2\times 10^{-10}$\,photons\,cm$^{-2}$\,s$^{-1}$.
Correspondingly, 
the TS value in time period II is 49, only slightly smaller than TS of 57 from 
the latter (data of totally 3701 days).
From the spectral analysis, we note that the two $\sim$478~GeV photon events 
determine whether the highest energy data point (273.9--500~GeV) in 
the spectrum in time period II is
a flux measurement or a flux upper limit (Figure~\ref{fig:spec}; see also 
Table~\ref{tab:spec}).

Given these, we conclude that there is approximately a
1.5-yr time period in which Tycho showed enhanced \gr\ emission in the 
energy range of 4--25\,GeV.
Based on the fluxes obtained from the binned likelihood analysis
to the data in time period II and the remaining time periods, the enhancement
has a significance of 3$\sigma$. However, our analyses involve the initial
search (from 5 light curves of from 100-day to 500-day bins) for possible 
signals
and determination of the optimal energy and time windows for the signal.
Although it is hard to evaluate a trial factor, 
i.e., the number of independent tests (e.g., \citealt{cho11}),
in the analyses, we may at least consider it to be much larger 
than 10, such as arising in the initial search and the search for the time 
range of the signal
(cf., Figure~\ref{fig:lc_4100}). Thus the significance should be lowered
to be less than 2$\sigma$.
Nevertheless, we did find evidence from the spectral analysis and TS maps,
plus the unusual two $\sim$478~GeV photons, indicating a possible enhancement
event in Tycho. Such an event would be the first time seen in a young SNR,
suggesting occurrences of complicated physical processes in young SNRs. 
Below
we discuss the possible models that might explain the enhancement and
their difficulties.


Tycho is young and there has been
discussion in regard to young SNRs noting that they may show \gr\ flux
variations due to the changeable ambient environments
their shock front goes through \citep{bkv11,bkv15,yl19}.
In particular, the number density of the target hadrons in the collision process
could have significant changes in the space surrounding a young SNR.
Previously on the observational side, X-ray hot spots or filaments in the young
SNRs RX~J1713.7$-$3946 and Cassiopeia~A were found to show yearly intensity
variations, which is interpreted as evidence for amplification of the magnetic
field (to $\sim$1~mG) at the shocks \citealt{uch+07,ua08}; but also
see \citealt{bud08}).
Also, the known
youngest SNR G1.9+0.3 ($\sim$100 yrs) in our Galaxy has been found to be
brightening at radio frequencies \citep{mgc08} and the remnant of
supernova~1987A (occurred in year 1987 in the Large Magellanic Cloud) was
brightening at both radio and X-ray frequencies \citep{zan+10,hel+13,fra+16}.
The phenomena either suggest magnetic field
amplification or the increasing volume and number of cosmic rays
\citep{bro+19} in the former, or reveal the interaction between the shock front
and the preexisting equatorial ring (formed from the progenitor system of
the supernova) in the latter.
In this respect, our results add a different, previously
unknown case to young SNRs by showing that a \gr\ emission
enhancement in a time period of more than a year can occur, in addition to
the baseline,
long-term emission variations. While the latter are understood
to reflect the physical processes SNRs go through \citep{rey17},
what this Tycho case reflects is intriguing. 

\subsection{\small Hadronic model}

\begin{figure*}
\centering
\includegraphics[width=0.56\linewidth]{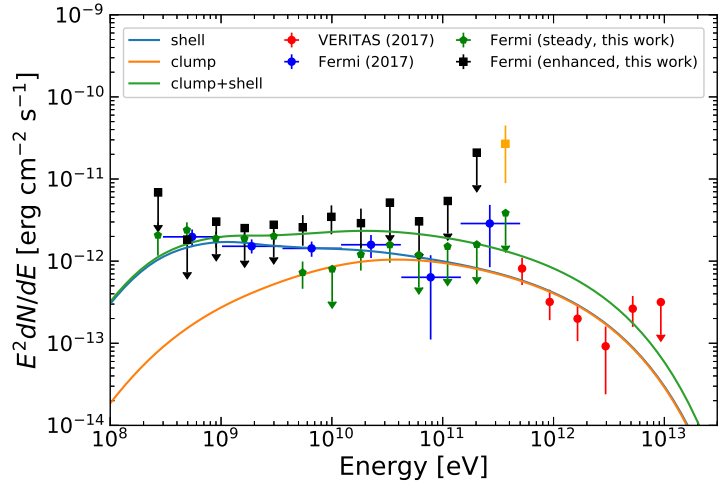}
\caption{Model fits to the \gr\ spectra of Tycho. The green pentagons
and black squares are the data points in the time period I plus III
(steady emission) and time period II (enhanced emission), respectively.
The highest-energy data point (orange square)
	that includes the two $\sim$478\,GeV photons
	is also shown.
For comparison,
the \gr\ spectra previously obtained with LAT and {\it Very Energetic
Radiation Imaging Array System (VERITAS)} \citep{arc+17} are shown as blue
and red dots, respectively.  A model spectrum from the shocked clump
in the hadronic process 
is shown as the orange curve, which can explain the enhanced emission when
it is added to the model spectrum (blue curve) for the steady emission.
The added spectrum is shown as the green curve.
\label{fig:sed}}
\end{figure*}

There must have been a temporary change of the high-energy emitting processes
related to Tycho in the time period around 2012 March to 2013 August
(MJD~56000--56530). The baseline (or steady) $\gamma$-ray emission of Tycho
is considered to arise from the hadronic
process \citep{mc12,zha+13,sla+14}.
The relativistic protons from the shock front have a long lifetime
$\tau_{\rm pp}=6\times10^{7}(n_{\rm t}/1\,{\rm cm}^{-3})^{-1}$~yr,
where $n_{\rm t}$ is the number density of the target gas.
Thus the $\gamma$-ray enhancement would have been only caused by
the change of the target, for which a probable scenario was that the shock front
swept up some dense clumps, and more protons were available as the target 
in the hadronic process. 
In such a case,
a transition layer around the boundary of a shocked clump is formed, which is
a highly turbulent region and will prevent low energy particles from
diffusively penetrating into the clump due to the amplification of
the magnetic field \citep{ino+12,ga14}.
Then a hard $\gamma$-ray spectrum (with a small spectral index)
arises from the shocked clump.
Applying this scenario, the increased $\gamma$-ray flux of Tycho in time
period II
could have been explained (Figure~\ref{fig:sed}).

Here we assume that the energetic protons accelerated by the shock take
away the explosion energy $E_{\rm sn}$ with a fraction of $\eta$ and
are uniformly distributed in the SNR shell. The shell is between
the forward shock (at radius $R_s$) and the contact discontinuity (at radius
$R_{\rm CD}$), and has width $\Delta R=R_s-R_{\rm CD}$, where
$R_{\rm CD} =rR_s$ and $r=0.93$ is adopted \citep{war+05}.
We consider that a dense clump with radius $R_c$ and density $n_c$
was swept up by the SNR shock at the beginning of time period II.
The $\gamma$-ray emission in this period is a sum of two components: one is
the emission from the SNR shell (or the steady component), and the other is
from the shocked clump.

The distribution in energy of the accelerated protons is
assumed to have a power-law form with a high-energy cutoff
\begin{equation}
dN/dE=N_0E_p^{-\alpha_p}\times \mathrm{exp}(-E_p/E_{c,p}),
\end{equation}
where $E_p$, $\alpha_p$, and $E_{c,p}$ are the proton energy,
the power-law index, and the cutoff energy, respectively. The normalization
$N_0$ is determined by the total energy in protons $\eta E_{\rm sn}$.
Modeling the steady component, the proton index $\alpha=2.2$, the ambient
density
$n_0 =0.1/(\eta/10\%)\,{\rm cm}^{-3}$, and the cutoff energy $E_{c, p}=20$~TeV
can be obtained (blue line in Figure~\ref{fig:sed}).
Given that the main differences between the spectra of the steady and the
enhanced emission are in the energy range of 4--20 GeV,
we take 10~GeV as the reference energy value.
To produce the enhanced component around 10 GeV, the protons with energies
above 100~GeV need to diffusively penetrate the transition layer and
reach the clump, which constrains the thickness of the transition layer
$L_{tr}=0.003$~pc and the magnetic field of it $B_{tr}=10$\,$\mu$G.
To fit the data, $R_c=0.05$ pc and
$n_c=1.2\times10^4/(\eta/10\%)\,{\rm cm}^{-3}$
are obtained (giving the orange line in Figure~\ref{fig:sed}).

This scenario seems to be
supported by the blast waves' reaching the wall of a surrounding
bubble \citep{zho+16} and rapid
deceleration of the SNR shell \citep{tan+20}, both of which likely indicate
the interaction of the blast waves with clumps of dense gas
and the latter notably occurred in a time period covering
the enhancement.
However, the time duration of a flux enhancement in this scenario should be
much longer than that of time period II, as
the shocked clumps in downstream of the shock have a survival time
described by the so-called cloud crushing timescale 
$\tau\approx 4\times 10^3(\chi/10^5)^{1/2}$\,yr 
($\chi=n_c/n_0 $; \citealt{kmc94}). We note that a similar scenario,
but with the cosmic-ray precursor interacting with the gas shell surrounding
an SNR, has been proposed by \citet{fed+15}, and it predicts a flux increase
at a few hundred GeV energies (which fits the enhancement event
in Tycho). Still the same problem exists, as the time duration should be
much longer.

\subsection{\small Inverse Compton scattering process}

The other often-considered $\gamma$-ray emission process is
inverse Compton (IC) scattering, in which
low-energy photons surrounding an SNR, consisting of the cosmic microwave
background (CMB), infrared background (IB), and star light (SL), are
boosted to the gamma-ray
domain by the relativistic electrons.
To produce photons of $\sim$10~GeV
energies via the IC scattering process, the electron energy needs to
be 1~TeV at least.
Due to the long lifetime of the TeV electrons ($\tau_{\rm IC}\sim10^6$~yrs),
the IC scattering process has difficulties generating year-timescale 
variations, unless the magnetic field could reach the order of milli-Gauss.
In such a strong magnetic field, the electrons will suffer severe
synchrotron energy loss, resulting in a sufficiently short lifetime
$\tau_{\rm syn}=1.4\ (E/1\ {\rm TeV})(B/3\ {\rm mG})^{-2}$ yr. Then
extremely high synchrotron emission (for example, with 10--20~keV flux of
$\sim10^{-8}\ \mathrm{erg\ cm^{-1}\ s^{-1} }$) will
be produced at the same time, which are several orders of magnitude higher than
the observed
($1.3\times 10^{-11}\ \mathrm{erg\ cm^{-1}\ s^{-1} }$; \citealt{tam+09}).
To avoid this inconsistency, one may reduce the density of
the relativistic electrons and increase that of the IB or SL photons.
However in order to explain the enhancement around 10~GeV emission, the energy
density of IB (SL) would need to be
$6000\ {\rm eV\,{\rm cm}^{-3}}$ ($30000\ {\rm eV\,{\rm cm}^{-3}})$,
which is unphysically high in the interstellar environment of Tycho.
Thus, the IC scattering process can be ruled out.

\subsection{\small Synchrotron emission modeling}

While high-energy electrons in the magnetic field can also produce
soft $\gamma$-rays via the synchrotron process, photon energies are limited
to be lower than the so-called synchrotron burnoff
limit ($\sim$160~MeV; \citealt{gfr83, dej+96}).
In order to exceed this limit,
the particle acceleration and cooling need to be decoupled.
A possible process that can provide additional relativistic electrons in
young SNRs is the magnetic reconnection (MR; \citealt{mat+15}). 
When the SNR shock sweeps up
some compact molecular clumps, the magnetic field around the interface between
the shock and clumps can be amplified \citep{ino+12}. The MR process may be
accompanied with the shock-clump interaction.
We also note that an MR detonation scenario has been specifically proposed 
for Tycho (and the Crab nebula), which actually suggests a year-long 
timescale for the reconnection process \citep{zgl18}.

Alternatively, the $\sim$10~GeV emission could be
ascribed to the synchrotron process in inhomogeneous magnetic fields.
The evidence for the magnetic field amplification in SNRs,
the variable X-ray structures (hot spots or filaments),
implies that the magnetic field in an SNR should have significantly different
strengths at different locations.
Specifically in Tycho, the X-ray stripes and their year-scale time variability
also indicate high inhomogeneity of the magnetic
field \citep{oku+20,eri+11,mat+20}.
In a recently discussed scenario for the presence of some discrete compact
magnetic blobs,
it has been shown that the maximum energy of synchrotron photons can be
boosted to multi-GeV energies depending on the ratio of the magnetic field
in the blobs to that in the background \citep{kha+20}.
For example, a synchrotron spectrum can extend to energies beyond 10\,GeV
for the ratio of $\sim$100.
If the $\gamma$-ray emitting magnetic blobs survive or decrease in the
field strength at yearly timescale, the $\gamma$-ray
enhancement might be explained.

However, in order to fit the enhanced part of the emission 
(cf., Figure~\ref{fig:sed}) with the synchrotron radiation, the electron
energies are required to be as high as $\sim 10$\,PeV (assuming a magnetic
strength of 1\,mG). Such high energies are above the ``knee" energy and
would have extremely short lifetimes of $\sim 1$\,day. It is hard to
explain the year-long enhancement event with such a short-lifetime radiation
process. 

In any case, we note that this possible $\gamma$-ray enhancement 
event in Tycho shares similar features with the $\gamma$-ray flares in 
the Crab nebula \citep{acrab,fcrab}.
Both occurred at GeV energies and their timescales are too short to be explained
by the processes other than the synchrotron. The complication in the Crab
nebula lies at the young pulsar in the center, which keeps powering
the nebula. For Tycho, it stands out among the young SNRs with the 
features of the variable X-ray strips,
blast-waves--bubble-wall interaction,
and rapid deceleration of the expansion. These features could be the reason
that induces the enhancement event, while detailed studies are needed in
order to fully understand the physical processes behind the event.


\acknowledgements
We thank anonymous referee for critical comments that helped shape up 
the manuscript, and J. Fang for helpful discussion about properties of pulsar wind nebulae and supernova remnants.
This research made use of the High Performance Computing Resource in the Core
Facility for Advanced Research Computing at Shanghai Astronomical Observatory. This work was supported by Key Research
and Development Project (Grant No. 2016YFA0400804) and the National Natural
Science Foundation of China (11633007, U1738131).
Z.W.  acknowledges the support by the Original
Innovation Program of the Chinese Academy of Sciences (E085021002) and
the Basic Research Program of Yunnan Province No. 202101AS070022.
X.Z. and Y.C. thank the support of National Key R\&D Program of China under 
nos. 2018YFA0404204 and 2017YFA0402600 and NSFC grants under nos. U1931204, 
12173018, and 12121003.

\clearpage
\begin{table}
\begin{center}
\setlength{\tabcolsep}{2pt}
\caption{Likelihood analysis results for Tycho}
\label{tab:gam}
\begin{tabular}{cccccccccccccccc}
\hline
\multicolumn{2}{c}{ } &
\multicolumn{3}{c}{0.3--500 GeV} &
\multicolumn{3}{c}{4.1--100 GeV} \\ \hline
Time & MJD & $\Gamma$ & $F/10^{-9}$ & TS & $\Gamma$ & $F/10^{-10}$ & TS \\
period & & & (photons cm$^{-2}$ s$^{-1}$) &  & & (photons cm$^{-2}$ s$^{-1}$) &
\\ \hline
Total & 54682--58913 & 2.23$\pm$0.07 & 4.3$\pm$0.5 & 245 & -- & -- & -- \\
I & 54682--56000 & 2.4$\pm$0.1 & 4.9$\pm$1.1 & 53 & -- & -- &
-- \\
II & 56000--56530 & 1.8$\pm$0.1 & 2.6$\pm$0.9 & 69 & 2.4$\pm$0.3 & 4.0$\pm$0.9 &
 49 \\
III & 56530--58913 & 2.31$\pm$0.09 & 5.0$\pm$0.7 & 140 & -- & -- & -- \\
I$+$III & -- & 2.35$\pm$0.08 & 4.9$\pm$0.6 & 192 & 1.9$\pm$0.3 & 1.1$\pm$0.2 & 57 \\
\hline
\end{tabular}
\end{center}
\end{table}

\begin{table}
\begin{center}
\setlength{\tabcolsep}{2pt}
\caption{Spectra obtained from the data in the whole time period, time period I
plus III, and time period II.}
\label{tab:spec}
\begin{tabular}{lccccccccccccc}
\hline
\multicolumn{2}{c}{ } &
\multicolumn{2}{c}{Whole data} &
\multicolumn{2}{c}{Period I$+$III} &
\multicolumn{2}{c}{Period II} \\ \hline
$E$ & Band & $F/10^{-12}$ & TS & $F/10^{-12}$ & TS & $F/10^{-12}$ & TS \\
(GeV) & (GeV) & (erg cm$^{-2}$ s$^{-1}$) &  & (erg cm$^{-2}$ s$^{-1}$) &  & (erg
 cm$^{-2}$ s$^{-1}$) & \\ \hline
0.27 & 0.2--0.4 & 2.3$\pm$0.8 & 12.8 & 2.1$\pm$0.9 & 9.1 & 6.8 & 4.4 \\
0.49 & 0.4--0.7 & 2.0$\pm$0.5 & 15.6 & 2.4$\pm$0.6 & 20.0 & 1.8 & 0.0 \\
0.90 & 0.7--1.2 & 1.8$\pm$0.4 & 21.3 & 1.9$\pm$0.4 & 20.5 & 3.0 & 1.2 \\
1.64 & 1.2--2.2 & 1.8$\pm$0.3 & 36.6 & 1.9$\pm$0.3 & 35.7 & 2.5 & 1.6 \\
3.00 & 2.2--4.1 & 1.9$\pm$0.3 & 58.6 & 2.0$\pm$0.3 & 56.3 & 2.8 & 3.0 \\
5.48 & 4.1--7.4 & 0.9$\pm$0.3 & 18.9 & 0.7$\pm$0.3 & 10.7 & 2.6$\pm$1.0 & 12.0 \\
10.00 & 7.4--13.5 & 0.7$\pm$0.3 & 15.1 & 0.8 & 3.5 & 3.4$\pm$1.3 & 22.0 \\
18.3 & 13.5--24.7 & 1.4$\pm$0.4 & 32.9 & 1.2$\pm$0.4 & 21.2 & 2.9$\pm$1.5 & 13.3
 \\
33.3 & 24.7--45.0 & 1.5$\pm$0.6 & 27.1 & 1.6$\pm$0.6 & 23.7 & 5.2 & 3.4 \\
60.8 & 45.0--82.2 & 1.0 & 1.4 & 1.2 & 1.7 & 3.0 & 0.0 \\
111.0 & 82.2--150.0 & 1.2 & 0.0 & 1.5 & 0.0 & 5.4 & 0.0 \\
202.7 & 150.0--273.9 & 1.3 & 0.0 & 1.6 & 0.0 & 20.9 & 0.0 \\
370.1 & 273.9--500.0 & 4.3$\pm$2.5 (3.3) & 16.7 (0.0) & 3.8 & 0.0 & 27$\pm$18 (18) & 23.9 (0) \\
\hline
\end{tabular}
\vskip 1mm
\footnotesize{Note: a) $F$ is the energy flux ($E^{2}dN/dE$), and fluxes without
 uncertainties are 95$\%$ upper limits. b) Values in parentheses were obtained
without including the two $\sim$478~GeV photons.}
\end{center}
\end{table}

\end{document}